\documentclass[twoside,12pt]{article}
\usepackage[dvips]{graphics}

\def\_{\lower 3pt\hbox{-}\nobreak}

\def\means{\mathrel{\!\hbox{\tt :-}}}
\def\commit{\mathrel{\hbox{\tt |}}}

\def\and{\mathpunct{\hbox{\tt ,}}}
\def\period{\mathpunct{\hbox{\tt .}}}
\def\equals{\!\mathrel{\hbox{\tt =}}\!}

\def\notni{\mathrel{\mathpalette\c@ncel\ni}}
\def\setminus{\mskip-4mu\mathrel{\backslash}\mskip-4mu}   
\def\definedas{\mathrel{\mathop{=}\limits^{\rm def}}}

\def\implies{\Rightarrow}

\def\qin{\mathbin{\in}}                

\def\pos#1#2{\langle {#1},\allowbreak {#2} \rangle}
\long\def\comment#1{}

\def\inmode{\hbox{\it in\/}}
\def\outmode{\hbox{\it out\/}}
\def\INmode{\hbox{\it IN\/}}
\def\OUTmode{\hbox{\it OUT\/}}
\def\Term{{\it Term}}
\def\Atom{{\it Atom}}
\def\singleout{{\cal R}}
\def\htt#1{\hbox{\tt #1}}
\def\hit#1{\hbox{\it #1}}

\def\.{\htt{.}}  
\def\unify{\htt{=}}
\def\HF{(HF)}
\def\HV{(HV)}
\def\GV{(GV)}
\def\BU{(BU)}
\def\BF{(BF)}
\def\BV{(BV)}

\newenvironment{FRAME}{\begin{trivlist}\item[]
	\hrule
	\hbox to \linewidth\bgroup
	\advance\linewidth by -20pt
	\hsize=\linewidth
	\vrule\hfill
	\vbox\bgroup
	\vskip10pt
	
	\begin{minipage}{\linewidth}}{
	\end{minipage}\vskip10pt
	\egroup\hfill\vrule
	\egroup\hrule
 \end{trivlist}}

\begin{document}
\pagestyle{myheadings}
\markboth{AADEBUG 2000}{Kima -- an Automated Error Correction System}
\title{Kima -- an Automated Error Correction System for Concurrent Logic
Programs
\footnote{In M. Ducass\'e (ed), proceedings of the Fourth International Workshop on Automated Debugging
(AADEBUG 2000), August 2000, Munich. COmputer Research Repository (http://www.acm.org/corr/),
cs.SE/0012007; whole proceedings: cs.SE/0010035.}}
\author{Yasuhiro Ajiro and Kazunori Ueda\\
Department of Information and Computer Science\\
Waseda University\\
{\tt \{ajiro,ueda\}@ueda.info.waseda.ac.jp}}
\date{}
\maketitle

\begin{abstract}
We have implemented Kima, an automated error correction system for
concurrent logic programs.  Kima corrects near-misses such as wrong
variable occurrences in the absence of explicit declarations of program
properties.

Strong moding/typing and constraint-based analysis are turning to play
fundamental roles in debugging concurrent logic programs as well as in
establishing the consistency of communication protocols and data types.
Mode/type analysis of Moded Flat GHC is a constraint satisfaction
problem with many simple mode/type constraints, and can be solved
efficiently.  We proposed a simple and efficient technique which, given
a non-well-moded/typed program, diagnoses the ``reasons'' of
inconsistency by finding minimal inconsistent subsets of mode/type
constraints.  Since each constraint keeps track of the symbol occurrence
in the program, a minimal subset also tells possible sources of program
errors.

Kima realizes automated correction by replacing symbol occurrences
around the possible sources and recalculating modes and types of
the rewritten programs systematically.  As long as bugs are near-misses,
Kima proposes a rather small number of alternatives that include an
intended program.  Search space is kept small because the minimal subset
confines possible sources of errors in advance.  This paper presents the
basic algorithm and various optimization techniques implemented in Kima,
and then discusses its effectiveness based on quantitative experiments.
\end{abstract}

\section{Introduction}

In our previous work \cite{CP98}, we proposed a framework of automated
debugging of program errors based on static, constraint-based program
analysis in the absence of programmers' declarations.  The framework was
then implemented in Kima, an automated error correction system for
concurrent logic programs, which featured several improvements to make
the system more practical and efficient.  

The mechanism of error correction in Kima is based on the mode and type
system of Moded Flat GHC \cite{ICLP90}\cite{NGC94}.  Moded Flat GHC is a
concurrent logic (and consequently, a concurrent constraint) language
with a constraint-based mode system designed by one of the authors.
%
Languages equipped with strong typing or strong moding\footnote{Modes
can be thought of as ``types in a broad sense,'' but in this paper we
reserve the term ``types'' to mean sets of possible values.} enable the
detection of type/mode errors by checking or reconstructing types or
modes.  The best-known framework for type reconstruction is the
Hindley-Milner type system \cite{Milner}, which allows us to solve a set
of type constraints obtained from program text efficiently as a
unification problem.

Similarly, the mode system of Moded Flat GHC allows us to solve a set of
mode constraints obtained from program text as a constraint satisfaction
problem.  Mode reconstruction statically determines the read/write
capabilities of variable occurrences and establishes the consistency of
communication protocols between concurrent processes \cite{NGC94}.
%
As we will see later, types in Moded Flat GHC also can be reconstructed
using a similar (and simpler) technique.


When a concurrent logic program contains bugs, it is very likely that
mode constraints obtained from the erroneous symbol occurrences are
incompatible with the other constraints.  We have proposed an efficient
algorithm that finds a minimal inconsistent subset of mode constraints
from an inconsistent (multi)set of constraints \cite{JICSLP96}.  
Since each
constraint keeps track of the symbol occurrence(s) in the program that
imposed the constraint, a minimal subset tells possible sources (i.e.,
symbol occurrences) of mode errors.

Using the information of possible locations of bugs, automated
correction is attempted basically by generate-and-test search, namely
the generation of possible rewritings and the computation of their
principal mode and type. Search space is kept small because the
locations of bugs have been limited to small regions of program text.

A significant feature of our framework is that it is applicable to a
fragment of a program such as a set of predicate definitions in a
particular module.  That is, our framework is quite effective, for
example, in the situation where a program is not completely constructed.
This is due to the fact that the multiset of mode constraints imposed
by a program usually has redundancy.
Redundancy comes from two reasons:
\begin{enumerate}
\item A non-trivial program contains conditional branches or
nondeterministic choices.  In (concurrent) logic languages, they are
expressed as a set of rewrite rules (i.e., program clauses) that may
impose the same mode constraints on the same predicate.

\item A non-trivial program contains predicates that are called from
more than one place, some of which may be recursive calls.  The same
mode constraint may be imposed by different calls.
\end{enumerate}
%

Although the framework is quite general, whether it is practical or not
may depend on the choice of a language.  
Kima corrects wrong occurrences of variable symbols in a KL1
\cite{Comp90} program assuming strong moding and typing of Moded Flat
GHC.  KL1 is designed based on Flat GHC that is not equipped with strong
moding/typing, but the debugging of KL1 programs turns out to benefit
from moding and typing.  Furthermore, its compiler KLIC provides a nice
platform for our experiments \cite{PLILP94}.
We have obtained promising results from our experiments with the
assistance of other syntactical constraints 
(Sect.~\ref{other-constraints}).

\section{Strong Moding and Typing in Concurrent Logic Programming}
\label{Strong-moding}

We outline the mode system of Moded Flat GHC.  The readers are referred
to \cite{NGC94} and \cite{PSLS95} for details.

In concurrent logic programming, modes play a fundamental
role in establishing the safety of a program in terms of the
consistency of communication protocols.  The mode system of Moded Flat
GHC gives a polarity structure (that determines the information flow
of each part of data structures created during execution) to the
arguments of predicates that determine the behavior of goals.  A mode
expresses this polarity structure, which is represented as a mapping
from the set of
{\em paths\/} to the two-valued codomain $\{\inmode,\outmode\}$.
Paths here are strings of pairs, of the form
$\pos{\hit{symbol}}{\hit{arg}}$, of predicate/function symbols and
argument positions, and are used to specify possible positions in data
structures.  Formally, the set $P_\Term$ of paths for terms and
the set $P_\Atom$ of paths for atomic formulae are defined using
disjoint union as:
$$P_\Term = (\sum_{f \in {\it Fun}}  N_f)^{*}\,,\;
  P_\Atom = (\sum_{p \in {\it Pred}} N_p)\times P_\Term\enspace\,,$$
where $\hit{Fun\/}$/$\hit{Pred\/}$ are the sets of function/predicate
symbols, and $N_f$/$N_p$ are the sets of possible argument positions
(numbered from 1) for the symbols $f$/$p$.  The disjoint union operator
$\sum$ means:
$$\sum_{f \in {\it Fun}} \!\!\!\! N_f = \{ \pos{f}{i}~|~f\in {\it Fun},
~i\in N_f \}\,\,.$$

The purpose of mode analysis is to find the set of all modes (each of
type $P_\Atom \rightarrow\{\inmode, \outmode\}$) under which every
piece of communication is cooperative.  Such a mode is called a {\em
well-moding}.  Intuitively, $\inmode$ means the inlet of information
and $\outmode$ means the outlet of information.
A program does not usually define a unique
well-moding but has many of them.  So the purpose of mode analysis is
to compute the set of all well-modings in the
form of a {\em principal\/} (i.e., most general) mode.  Principal modes
can be expressed naturally by mode graphs, as described later in this
section.

Given a mode $m$, we define a {\em submode\/} $m/p$, namely $m$
viewed at the path $p$, as a function satisfying $(m/p)(q) = m(pq)$.
We also define $\INmode$ and $\OUTmode$ as submodes returning
$\inmode$ and $\outmode$, respectively, for any path.  An overline
`$\,\overline
{\hbox{\vrule height4pt depth3.5pt width0pt}~~}\,$' inverts the
polarity of a mode, a submode, or a mode value.

\begin{figure}
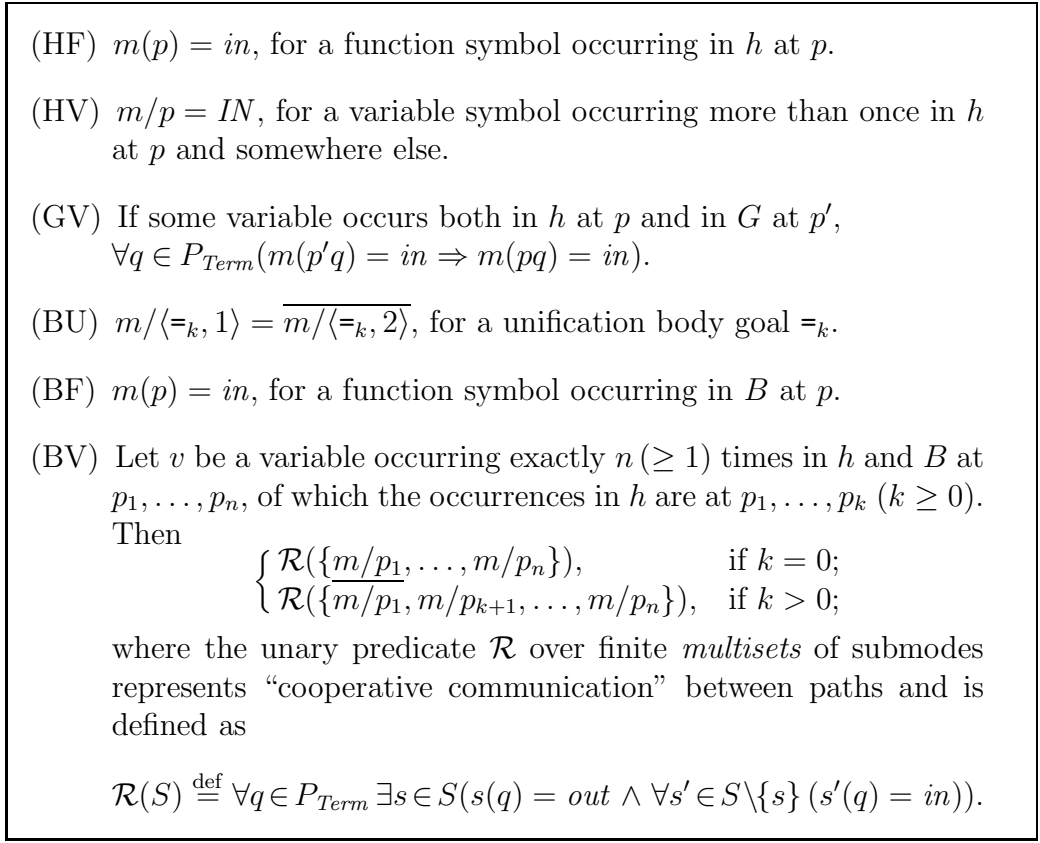

\begin{FRAME}
\begin{list}{}{\setlength{\leftmargin}{30pt}\setlength{\labelwidth}{25pt}
\setlength{\itemsep}{\medskipamount}\setlength{\rightmargin}{10pt}
}
\item[\HF]
$m(p)=\inmode$, for a function symbol occurring in $h$ at $p$.

\item[\HV]
$m/p = \INmode$, for a variable symbol occurring more than once in $h$ at
$p$ and somewhere else.

\item[\GV]
If some variable occurs both in $h$ at $p$ and in $G$ at $p'$,\\
$\forall q\qin P_\Term \bigl(m(p'q)=\inmode \implies m(pq)=\inmode\bigr)$.

\item[\BU]
$m/\pos{\unify_k}{1} = \overline{m/\pos{\unify_k}{2}}$,
for a unification body goal $\unify_k$.

\item[\BF]
$m(p)=\inmode$, for a function symbol occurring in $B$ at $p$.

\item[\BV]
Let $v$ be a variable occurring exactly $n\, (\ge 1)$ times in $h$ and
$B$ at $p_1, \ldots, p_n$, of which the occurrences in $h$ are at $p_1,
\ldots, p_k$ ($k\ge 0$).  Then
$$\cases{
\singleout\bigl(\{ m/p_1, \ldots, m/p_n\}\bigr),&if $k=0$;\cr
\singleout\bigl(\{ \overline{m/p_1}, m/p_{k+1}, \ldots, m/p_n\}\bigr),&
   if $k>0$;\cr}$$
where the unary predicate $\singleout$ over finite {\em multisets} of submodes
represents ``cooperative communication'' between paths and is defined as
$$\singleout(S)\definedas 
  \forall q\qin P_\Term\,\exists s\qin S
    \bigl(s(q)=\outmode \,\land\,
           \forall s'\qin S\setminus\!\{s\}\,\bigl(s'(q)=\inmode\bigr)\bigr).
$$
\end{list}
\end{FRAME}
\caption{Mode constraints imposed by a program clause $h\means G\commit B$
or a goal clause $\means B$.}
\label{ModeConstraints}
\end{figure}

A Flat GHC program is a set of clauses of the form $h\means G\commit B$,
where $h$ is an atomic formula and $G$ and $B$ are multisets of atomic
formulae.  Mode constraints imposed by a clause $h\means G\commit B$ are
summarized in Fig.~\ref{ModeConstraints}.  All rules here embody the
assumption that every piece of communication is cooperative.
%
Rule \BU\ numbers unification body goals because the mode system
allows different body unification goals to have different modes.  This
is a special case of mode polymorphism that can be introduced into other
predicates as well \cite{JICSLP96}, but in this paper we will not
consider general mode polymorphism because whether to have
polymorphism is independent of the essence of this work.

\begin{figure}
\begin{FRAME}
\begin{center}
\scalebox{0.5}{\includegraphics{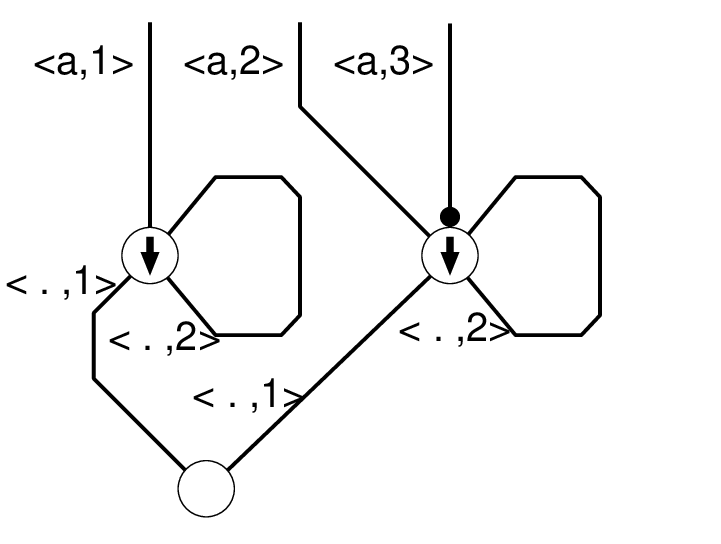}}
\end{center}
\end{FRAME}
\caption{The mode graph of an append program.  ``$\htt{a}$'' stands for
$\htt{append}$; ``$\htt{\.}$'' stands for list constructor; and the
downward arrow means the mode value $\inmode$.  The mode information of
the toplevel predicate and unification goals is omitted.}
\label{graph:append}
\end{figure}

The cost of mode analysis is almost proportional to the program size for
the following reason.  Mode analysis proceeds by merging many simple
mode graphs representing individual mode constraints.  For example, the
resulting mode graph of the append program (cf. Appendix) is shown in
Fig.~\ref{graph:append}.  The mode graphs of very large programs are, in
general, much wider than that of the append program but are not much
deeper, which is to say most nodes can be reachable within several steps
from the root.  The cost of merging one mode constraint with a mode graph
is almost proportional to the depth of the mode graph, but does not
depend on the width of the graph \cite{PSLS95}.  So the total cost is
proportional to the number of constraints that in turn is proportional
to the program size.

%
%
%

A type system for concurrent logic programming can be introduced by
classifying the set $\hit{Fun\/}$ of function symbols into mutually
disjoint sets $F_1,\ldots,F_n$.  A type here is a function from
$P_\Atom$ to the set $\{F_1,\ldots,F_n\}$.  Like principal modes, 
principal types can be computed by unification over feature graphs.
Constraints on a well-typing $\tau$ are summarized in Fig.~
\ref{TypeConstraints}.
The choice of a family of sets $F_1,\ldots,F_n$ is somewhat arbitrary.
This is why moding is more fundamental than typing in concurrent logic
programming.


\begin{figure*}
\begin{FRAME}
\begin{list}{}{\setlength{\leftmargin}{40pt}\setlength{\labelwidth}{35pt}
\setlength{\itemsep}{\medskipamount}\setlength{\rightmargin}{10pt}}
\item[$({\rm HBF}_\tau)$]
$\tau(p)=F_i$, for a function symbol occurring at $p$ in $h$ or $B$.

\item[$({\rm HBV}_\tau)$]
$\tau/p=\tau/p'$, for a variable occurring both at $p$ and $p'$ in 
$h$ or $B$.

\item[$({\rm GV}_\tau)$]
$\forall q\qin P_\Term \bigl(m(p'q)=\inmode \implies
                           \tau(pq)=\tau(p'q)\bigr)$, for a variable
occurring both at $p$ in $h$ and at $p'$ in $G$.

\item[$({\rm BU}_\tau)$]
$\tau/\pos{\unify_k}{1} = \tau/\pos{\unify_k}{2}$, 
for a unification body goal $\unify_k$.
\end{list}
\end{FRAME}
\vskip-6pt
\caption{Type constraints imposed by a program clause $h\means
G\commit B$ or a goal clause $\means B$.}
\label{TypeConstraints}
\end{figure*}

The type system employed by Kima classifies function symbols into six
disjoint sets --- integers, floating-point numbers, strings, vectors,
lists and functor structures, and prohibits any two of them from sharing
the same path.  Although this is a heuristic classification based on the
fact that these different types do not simultaneously appear in the same
path in most programs, an experiment proves that it is beneficial both
to the power of error detection and to the quality of error correction,
as we will see in Sect.~\ref{experiment}.

\section{Identifying Program Errors}
\label{Identification}

When a concurrent logic program contains an error, it is very likely
(though not always the case) that its communication protocols become
inconsistent and the set of its mode constraints becomes unsatisfiable.
A wrong symbol occurring at some path is likely to impose a mode
constraint inconsistent with constraints representing the intended
specification.

Then, suspicious symbols can be located by computing a minimal
inconsistent subset of mode constraints, because the minimal
inconsistent subset must include at least one wrong constraint, and each
constraint is imposed on certain symbol occurrences in a clause
(see the moding rules in Fig.~\ref{ModeConstraints}).  Type constraints
can be used in the same way to locate type errors.

A minimal inconsistent subset can be computed efficiently using a simple
algorithm shown in Fig.~\ref{algorithm-mis1}
\footnote{The algorithm described
here is a revised version of the one proposed in \cite{JICSLP96} and
takes into account the case when $C$ is consistent.}.
Let $C=\{c_1,\dots,c_n\}$ be a multiset of constraints.  The algorithm
finds a single minimal inconsistent subset $S$ from $C$ when $C$
is inconsistent.  When $C$ is consistent, the algorithm terminates with
$S=\{\}$.  $false$ is a self-inconsistent constraint used as a sentinel.

\medskip
\begin{figure}
\begin{FRAME}
\begin{list}{}{%
\setlength{\leftmargin}{40pt}%
\setlength{\topsep}{0pt}%
\setlength{\partopsep}{0pt}%
\setlength{\itemsep}{0pt}%
\setlength{\parsep}{0pt}}
\item $ c_{n+1} \leftarrow false;\; $
\item $ S \leftarrow \{\}; $
\item {\bf while} $S$ is consistent {\bf do}
\item \ \ $ D \leftarrow S ;\; i \leftarrow 0; $
\item \ \ {\bf while} $D$ is consistent {\bf do}
\item \ \ \ \ $ i \leftarrow i+1;\;
                D \leftarrow D \cup \{c_i\} $
\item \ \ {\bf end while};
\item \ \ $ S \leftarrow S \cup \{c_i\} $
\item {\bf end while};
\item {\bf if} $i=n+1$ {\bf then} $S \leftarrow \{\}$ {\bf fi}
\end{list}
\end{FRAME}
\caption{Algorithm for computing a minimal inconsistent subset}
\label{algorithm-mis1}
\end{figure}

The readers are referred to \cite{JICSLP96} for a proof of the
minimality of $S$, as well as various extensions of the algorithm.
Note that the algorithm can be readily extended to finding multiple
bugs at once.  That is, once we have found a minimal subset covering a 
bug, we can reapply the algorithm to the rest of the constraints.


Our experiment shows that the average size of minimal inconsistent
subsets is 
rather small, and the subsets containing more than 10 elements are
scarcely found.  The size of minimal subsets turns out to be independent
of the total number of constraints, and most inconsistencies can be
explained by constraints imposed by a small range of program text.  This
is due to the redundancy of mode and type constraints.


\section{Automated Debugging}
\label{Automated-debugging}

Constraints that are considered wrong may be corrected by
\begin{itemize}
\item replacing the symbol occurrences that imposed those constraints by
      other symbols, or 
\item when the suspected symbols are variables, by making them have more
occurrences elsewhere (cf. Rule \BV\ of Fig.~\ref{ModeConstraints}).
\end{itemize}
When some symbol occurrence has been rewritten to another symbol by
mistake, there exists a symbol with less occurrences than intended and a
symbol with more occurrences.  A minimal inconsistent subset includes
either (or both) of them.

Kima focuses on programs with a small number of errors in variables.
This may sound restrictive, but concurrent logic programs have quite
flat syntactic structures (compared with other languages) and instead
make heavy use of variables.  Our experience tells that a majority of
simple program errors arises from the erroneous use of variables, for
which the support of a static mode and type system and debugging tools
are invaluable.

This technique is applicable also to mutations between a constant and a
variable symbol, because mode and type constraints are imposed also on
constant symbols.  Even mutations between constant symbols could be
corrected by type constraints.  However, when considering replacement by
a constant symbol, Kima must determine its value.  It is difficult for
the current version of Kima to determine the value based on modes and
types only.  Mutations of function symbols (other than constant symbols)
can also be located but their correction is difficult because search
space will expand too much.  Mutations of predicate symbols cannot be
corrected by the current framework.

\subsection{Basic Algorithm}
\label{algorithm-correct}

An algorithm for automated correction is basically a search procedure
whose initial state is the erroneous program, whose operations are the
rewriting of the occurrences of variables, and whose final states are
well-moded/typed programs.



The algorithm in Fig.~\ref{algorithm-kima} finds a set $S$ of
alternative solutions.  The main procedure of the algorithm is
iterative-deepening search up to the maximum depth MAX which is to be
given by a user.  $\delta$ represents the current depth.

\begin{figure}
\begin{FRAME}
\begin{list}{}{%
\setlength{\leftmargin}{10pt}%
\setlength{\topsep}{0pt}%
\setlength{\partopsep}{0pt}%
\setlength{\itemsep}{0pt}%
\setlength{\parsep}{0pt}
}
\item Compute a minimal inconsistent subset of mode/type constraints;
\item Extract suspicious variable symbols from the subset;
\item $ \delta \leftarrow 1; \;\; S \leftarrow \{\};$
\item {\bf while} MAX $\geq \delta$ {\bf do}
\item \ \ {\bf for each} way of rewriting $\delta$ symbol occurrences {\bf do}
\item \ \ \ \ {\bf if} the rewritten program becomes well-moded/typed
{\bf then} 
\item \ \ \ \ \ \ Add the way of rewriting to $S$ {\bf fi}
\item \ \ {\bf end for};
\item \ \ $\delta \leftarrow \delta +1$
\item {\bf end while}
\end{list}
\end{FRAME}
\caption{The basic algorithm for automated error correction}
\label{algorithm-kima}
\end{figure}

%

Since checking modes and types of a rewritten program requires the cost
proportional to the program size (Sect.~\ref{Strong-moding}), this
algorithm takes time in proportion to the program size.  However,
inconsistency usually occurs within a small region of program text
(Sect.~\ref{Identification}).  A large performance improvement will
therefore be achieved by analyzing those constraints imposed by the
suspected predicates and its closely related predicates before the whole
constraints.

\subsection{Grouping Errors}
\label{grouping}

\begin{figure}
\begin{center}
\scalebox{0.7}{\includegraphics{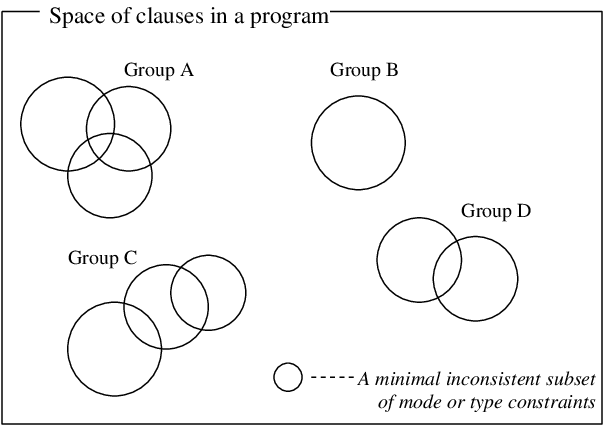}}
\end{center}
\vspace{-12pt}
\caption{Grouping minimal inconsistent subsets}
\label{grouping-mis}
\end{figure}

As we mentioned in Sect.~\ref{Identification}, multiple minimal
inconsistent subsets may independently be found, and some of them may
indicate the same clause as the source of errors.  The clause may be
indicated by subsets of modes, types, or both.  Modes and types express
different properties of a program and detect different kinds of errors.
To use them together makes two improvements; one is that more errors can
be detected; the other is that errors can be located more precisely.
Kima collects minimal inconsistent subsets indicating the same clause
(as in Fig.~\ref{grouping-mis}), and makes them belong to the same group
which plays the role of a unit of searching alternatives against errors.


Depth-$\delta$ search of alternatives is carried out independently
for each group.  It is possible to reduce computation time by checking
whether a certain rewriting can possibly dissolve inconsistencies of all
minimal inconsistent subsets in a group before actually computing modes
and types, which is called {\em Quick-check}.
This is effective because when some symbol occurrence is rewritten, even
if the change is small, mode and type analyses may reanalyse the whole
program.
%
To put it more precisely, Quick-check of a rewriting is to check if, for
all minimal inconsistent subsets in a group, there exists a variable
indicated as a possible source of mode/type errors such that the given
rewriting will result in:
\begin{itemize}
\item replacing an occurrence of the indicated variable by a different
      variable, or
\item making more occurrences of the indicated variable elsewhere.
\end{itemize}

When multiple groups are found, mode and type analyses are performed
with the constraints imposed by one of the groups and consistent part.
The consistent part is the set of all constraints which do not belong to
any minimal inconsistent subset.  Therefore, not all constraints imposed
by the whole program text are considered in error correction.  Kima employs
such an algorithm so that the search of alternatives for one group may
not be influenced by that for another group.  
Alternatives found for a group do not always include an intended one.

%

\section{Constraints Other Than Modes and Types}
\label{other-constraints}


\subsection{Prioritizing Alternatives}
\label{prioritize}

Kima searches alternative solutions using mode and type information, but
unfortunately, multiple alternatives are found in many cases.  Kima
refines the quality of its output by prioritizing alternatives using two
{\em Heuristic Rules}:
\begin{description}
 \item[Heuristic Rule 1.] It is less likely that a variable occurs
\begin{enumerate}
 \item only once in a clause (singleton occurrence),
 \item two or more times in a clause head,
 \item three or more times in the head and/or the body of a clause, or
 \item two or more times as arguments of the same body goal.
\end{enumerate}
 \item[Heuristic Rule 2.] It is less likely that a list and its elements
	    are of the same type, that is, it is less likely that a
	    variable occurs both in some path $p$ and in the path of its
	    elements $p\pos{\.}{1}$.
\end{description}
%

Since such variable occurrences as in Heuristic Rules 1.1, 1.2 and 1.3
impose mode constraints \INmode \/ or \OUTmode \/
(Sect.~\ref{Strong-moding}) that are stroger than \inmode \/ and
\outmode, we could replace Heuristic Rules 1.1-1.3 by a unified rule on
constraint strength; {\it A solution with weaker mode constraints is
more likely to be an intended one}.
In general, stronger mode/type constraints make a program less generic,
and the execution of the program more likely to end in failure.
Therefore it is reasonable to insist that the constraint imposed on a
program should be as weak as possible.

Heuristic Rules 1.1 and 1.3 are justified also in the sense that logical
variables are used for one-to-one communication more frequently than for
one-to-many or one-to-zero communication.  A logical variable used for
one-to-one communication occurs either exactly twice in the clause
body or exactly once in the clause head and once in the clause body.
Such a body goal as in Heuristic Rule 1.4 either receives duplicate data
from another goal or communicates with itself, which are both unlikely.

For Heuristic Rule 2, let $\alpha$ be a type variable and $list(\alpha)$
be the list type whose elements are of type $\alpha$. Then the rule is
equivalent to saying that constraint $\alpha = list(\alpha )$ imposes a
strong type constraint on $\alpha$ and is therefore unlikely.

Kima prioritizes multiple alternatives by imposing certain penalty
points on unlikely symbol occurrences.  An alternative with a lower
penalty point has a higher priority.

\subsection{Reinforcing Detection Power}
\label{detection-level}

The objective of Kima is to debug a program in the absence of explicit
declarations.  To enhance the power of the error detection with implicit
modes and types, Kima employed the following {\em Detection Rules}:
\begin{description}
 \item[Detection Rule 1.] \hspace{10cm}
\begin{enumerate}
 \item A variable which occurs in a clause guard must occur also in the
       head of the clause.
 \item The same variable must not occur on both sides of a unification
       body goal (partial occur-check).
\end{enumerate}
 \item[Detection Rule 2.]
The name of a singleton variable must begin with an underscore ``{\tt \_}''.
\end{description}
The both Detection Rules are optional and can be used selectively in Kima.

Violation of Detection Rule 1.1 means disappearance of a guard goal
after normalization \cite{NGC94}, while violation of Detection Rule 1.2
means failure of normalization.
%
Detection Rule 2 is identical to requesting the declaration of variables
that impose strong mode constraints.
Detection Rule 2 is effective because a logical variable in a correct
program is likely to occur twice in a clause (i.e., for one-to-one
communication), in which case a variable will turn into a singleton if
either occurrence is missing.

The source of an error detected by Detection Rules is a variable symbol
in a certain clause, and is found independently of minimal inconsistent
subsets of mode and type constraints.  Kima uniformly deals with a
variable symbol detected by Detection Rules by considering it as a
minimal inconsistent subset with one element, and groups it with other
subsets.

\subsection{Optimizing Search of Alternatives}
\label{optimize}

Kima employs two optimization techniques other than Quick-check.  The
two techniques are based on prioritizing and Detection Rules stated in
Sect.~\ref{prioritize} and \ref{detection-level}, and reduce the number
of mode and type analyses in generate-and-test search.
The algorithm shown in Fig.~\ref{opt-algorithm} finds a set $S$ of
alternatives that have higher priorities than the given priority $P$.
Steps related to grouping process (Sect.~\ref{grouping}) are omitted.

\medskip
\begin{figure}
\begin{FRAME}
\begin{list}{}{%
\setlength{\leftmargin}{15pt}%
\setlength{\topsep}{0pt}%
\setlength{\partopsep}{0pt}%
\setlength{\itemsep}{0pt}%
\setlength{\parsep}{0pt}
}
\item Compute a minimal inconsistent subset of mode/type constraints;
\item Extract suspicious variable symbols from the subset;
\item Detect clauses and variable symbols infringing Detection Rules;
\item $ \delta \leftarrow 1; \;\; S \leftarrow \{\};$
\item {\bf while} MAX $\geq \delta$ {\bf do}
\item \ \ {\bf for each} way of rewriting $\delta$ symbol occurrences
		which has 
\item \ \ \ \ \ \ \ \ \ \ \ \ \ \ \ a higher priority than $P$
		w.r.t. Heuristic Rule 1 {\bf do}
\item \ \ \ \ {\bf if} the rewritten program follows Detection Rules {\bf then}
\item \ \ \ \ \ \ {\bf if} the rewritten program becomes
		well-moded/typed {\bf then}
\item \ \ \ \ \ \ \ \ {\bf if} the rewriting is unlikely
		w.r.t. Heuristic Rule 2 {\bf then}
\item \ \ \ \ \ \ \ \ \ \ lower the priority of the rewriting
\item \ \ \ \ \ \ \ \ {\bf fi};
\item \ \ \ \ \ \ \ \ Add the way of rewriting to $S$ with its priority
\item \ \ \ \ \ \ {\bf fi}
\item \ \ \ \ {\bf fi}
\item \ \ {\bf end for};
\item \ \ $ \delta \leftarrow \delta +1$
\item {\bf end while}
\end{list}
\end{FRAME}
\caption{The optimized algorithm for automated error correction}
\label{opt-algorithm}
\end{figure}

In generate-and-test search, the test by Detection Rules is cheaper than
mode and type analyses, because, when particular (suspected) clauses are
rewritten, the clauses that are not rewritten do not have to be checked
with Detection Rules again.  In contrast, mode and type analyses may need
recalculation of the whole program (Sect.~\ref{grouping}).

Prioritizing with Heuristic Rule 1 involves only suspected clauses, and
is cheaper than mode and type analyses.
%
Prioritizing with Heuristic Rule 2 needs type analysis, and is performed
after the check of well-typedness (i.e., type reconstruction).

For a quicksort program containing two wrong variable occurrences in the
same clause (Example 3 in Appendix), the above optimization improved the
response time of computing highest-priority alternatives from 25.9
seconds to 10.2 seconds on the KLIC system running on Sun Ultra 30 (248
MHz) + 128 MB of memory.

\section{Experiments}
\label{experiment}

We discuss the effectiveness of our technique based on experiments.  We
investigated how many of programs with a few errors are detected as
erroneous by Kima, how many alternatives it proposes for erroneous
programs, and how many ``plausible'' programs there are in the
neighborhood of a correct (original) program.

%
First, we systematically generated near-misses (each with one wrong
occurrence of a variable) of three programs and examined how many of
them could be detected, whether automated correction reported an
intended program, and how many alternatives were reported.
Table \ref{Result-kima} shows the results
\footnote{In a similar experiment shown in our previous paper
\cite{CP98}, the numbers are different because errors in the clause
guard and those concerning Detection Rule 1 were not counted and errors
detected by types but not detected by modes were not considered by
automated debugging.}.
Here, we did not consider the mutation of a variable occurrence to the
variable whose name began with ``{\tt \_}''.  We used only the
definitions of predicates, that is, we did not use the constraints that
might be imposed by the caller of these programs.  Of course, the caller
information, if available, would enhance the quality of correction as
well as the redundancy of constraints.

\begin{table}
\caption{Single-error detection and correction}
\label{Result-kima}
\vspace{-6pt}
{\small
\begin{center}
\begin{tabular}{c@{~}c@{~}c@{~}c@{~}r@{~}r@{~~~}r@{~~}r@{~~}r@{~~}r@{~~}r@{~~}r@{~~}r@{~~}}
\hline\noalign{\vskip2pt}
Program& Analysis&Level&Priori-&Total&Dete-&\multicolumn{7}{c}{Proposed Alternatives}\\[-1pt]
       &         &     & tizing&cases& cted&~1&~2&~3&~4&~5&~6&$\mathord{\ge}7$\\
\hline\noalign{\vskip2pt}
{\tt append}    &mode only   & 0&  no&  58&  36&  1&  3& 8& 3& 6& 5& 10\\
                &type only   & 0&  no&  58&   0&  0&  0& 0& 0& 0& 0&  0\\
                &mode \& type& 0&  no&  58&  36&  1&  3& 8& 3& 6& 5& 10\\
                &mode \& type& 0& yes&  58&  36& 27&  9& 0& 0& 0& 0&  0\\
                &mode \& type& 1& yes&  58&  40& 29& 11& 0& 0& 0& 0&  0\\
                &mode \& type& 2& yes&  58&  58& 39& 19& 0& 0& 0& 0&  0\\ \hline
\noalign{\vskip2pt}
{\tt fibonacci} &mode only &   0&  no& 118&  57& 10&  7&  9&  6& 4&  1& 20\\
                &type only &   0&  no&  118&  47&  0&  0&  4& 20& 0& 18&  5\\
                &mode \& type& 0&  no& 118& 72& 18& 13&  2& 15& 9& 0& 15\\
                &mode \& type& 0& yes& 118&  72& 54& 11&  1&  6& 0&  0&  0\\
                &mode \& type& 1& yes& 118&  88& 68& 12&  7&  0& 1&  0&  0\\
                &mode \& type& 2& yes& 118&  99& 71& 18&  8&  0& 2&  0&  0\\ \hline
\noalign{\vskip2pt}
{\tt quicksort} &mode only &   0&  no& 300& 177& 34& 70&  1& 12& 19& 0& 41\\
                &type only &   0&  no& 300& 106&  0&  2& 12& 40&  0&32& 20\\
                &mode \& type& 0& no&  300& 221& 49& 76&  8& 59&  0& 9& 20\\
                &mode \& type& 0& yes& 300& 221&164& 41& 16&  0&  0&  0& 0\\
                &mode \& type& 1& yes& 300&  236&175& 61&  0&  0&  0& 0&  0\\
                &mode \& type& 2& yes& 300&  286&199& 84&  2&  1&  0& 0&  0\\ \hline
\end{tabular} 
\end{center}
}
\end{table}

The programs we used are list concatenation (append), the generator of a
Fibonacci sequence, and quicksort.
%
%
%
%
%
%
%
They are admittedly simple but the aim of the experiment is to investigate
the fundamental power of our technique based on exhaustive experiments.
Further, for the reason discussed in Sect.~\ref{algorithm-correct},
we strongly expect that the total program size does not make much
difference in the quality of automated debugging.

The column ``Level'' indicates detection levels.
Under detection level 0, only mode and/or type information was used;
under detection level 1, Detection Rule 1 was used in addition; and
under detection level 2, Detection Rules 1 and 2 were used together.
The two Detection Rules raised the average detection rate from 
69.1\% (329/476) to  93.1\% (443/476).

A row with ``yes'' in the column ``Prioritizing'' shows the number of
proposed alternatives with the highest priority, which includes an
intended alternative in most cases.  The number of proposed alternatives
under prioritizing was usually 1 or quite small.
%

%
Second, we investigated the error detection rate for programs with two
and three mutated variable occurrences in the same clause.  Errors of
this kind are looked on as depth-2 and depth-3 errors in the same group,
respectively, and their correct alternatives can be obtained by depth-2
and depth-3 search.  Table \ref{dist-consistent} shows the results.
Note that the mutation of variable occurrences does not always cause
errors.  For example, certain mutations make a program equivalent to the
original as we will see later.

\begin{table}[tb]
\caption{The error detection rate for the programs with N mutations}
\label{dist-consistent}
\vspace{-6pt}
\begin{center}
\begin{tabular}{c@{\quad}c@{~~}c@{~~}r@{~~}@{\quad}r@{~~}r}
\hline
Program         & N&  Level&  Total& Detected& Detection \\ 
                &  &       &  cases&    cases&   rate (\%)\\ \hline
{\tt append}    & 2&      0&   1200&    937&  78.1\\
                & 2&      1&   1200&   1004&  83.7\\
                & 2&      2&   1200&   1141&  95.1\\
                & 3&      0&  16980&  14597&  86.0\\
                & 3&      1&  16980&  15411&  90.8\\
                & 3&      2&  16980&  16674&  98.2\\ \hline
{\tt fibonacci} & 2&      0&   4668&   3982&  85.3\\
                & 2&      1&   4668&   4330&  92.8\\
                & 2&      2&   4668&   4489&  96.2\\
                & 3&      0& 133045& 125300&  94.2\\
                & 3&      1& 133045& 130325&  97.9\\
                & 3&      2& 133045& 131810&  99.1\\ \hline
{\tt quicksort} & 2&      0&  12102&  11263&  93.1\\
                & 2&      1&  12102&  11460&  94.7\\
                & 2&      2&  12102&  12005&  99.2\\
                & 3&      0& 337455& 330769&  98.0\\
                & 3&      1& 337455& 332416&  98.5\\
                & 3&      2& 337455& 336943&  99.8\\ \hline
\end{tabular} 
\end{center}
\end{table}

When multiple errors existed in some clause of a program, the program
was detected as long as at least one of the errors caused inconsistency.
So the detection rate for multiple errors was higher than that for a
single error.  The detection rate with Detection Rules 1 and 2 was above
95\% in every case.

%
Third, we explored the number of ``plausible'' programs.  Plausible
programs are programs that have the same or higher priority than the
original among the programs with N mutations on variable occurrences in
the same clause for a certain N.  The result is shown in
Table~\ref{dist-highpriority}.  We considered the mutations to the
variable whose name began with ``{\tt \_}''.  In the column ``Plausible
programs'', programs that were
\begin{enumerate}
 \item equivalent up to renaming of variables and
 \item equivalent up to switching of arguments of calls to commutative
       built-in predicates such as unification
\end{enumerate}
were counted as one program.

\begin{table}[tb]
\caption{The number of plausible programs among the programs with N mutations}
\label{dist-highpriority}
\vspace{-6pt}
\begin{center}
\begin{tabular}{c@{\quad}r@{~~~~~}r@{~~}r}
\hline
Program      & N& Total& Plausible\\
             &  & cases& programs\\ \hline
{\tt append}    & 1&       58&   0\\
                & 2&     1200&   7\\
                & 3&    16980&  14\\
                & 4&   167842&  29\\ \hline
{\tt fibonacci} & 1&      118&  11\\
                & 2&     4668&  66\\
                & 3&   133045& 309\\ \hline
{\tt quicksort} & 1&      300&   9\\
                & 2&    12102&  33\\
                & 3&   337455&  76\\ \hline
\end{tabular} 
\end{center}
\end{table}

From Table~\ref{dist-highpriority} we see that the number of plausible
programs did not increase explosively.
This can be explained by the fact that the ways of placing variable
symbols which make a program well-moded/typed are extremely limited
compared to arbitrary ways of placing.

Now we focus on the number of proposed alternatives under prioritizing.
Suppose, for example, a program contains two errors on variable
occurrences and depth-2 search is performed.  In this case, up to four
occurrences may be rewritten from the original, in which the rewritings
with N=4 will be the majority.  However, since two of the four
rewritings have already been done by the given erroneous program, only
part of the rewritings where N is up to 4 is generated.  The total
number of rewritings generated actually is very close to the number of
rewritings with N=2.

Among the plausible programs, the percentage of programs that neither
diverge or fail depends on the original program and its expected input.
In the case of quicksort, about fifty percent of plausible programs were
programs that neither diverge or fail.  We note that, of these programs,
few were considered meaningful, that is, few programs were such that all
operations contribute to the result of computation.

\section{An Example --- {\tt Fibonacci sequence}}
\label{exam-fibonacci}

As an example, we consider the generator of a Fibonacci sequence with
one error:

\begin{FRAME}
\hbox{1:~~$R_1$ : $\htt{fib(Max,\_,~~N2,Ns0)}\means \htt{N2 >Max}\commit
\htt{Ns0} \equals_1 \htt{[]}\period$}
\hbox{2:~~$R_2$ : $\htt{fib(Max,N1,N2,Ns0)}\means \htt{N2=<Max}\commit$}
\hbox{3:~~~~~~~~~~~~$\htt{N1}\equals_2\htt{[N2|Ns1]}\and \htt{N3:=N1+N2}\and
\htt{fib(Max,N2,N3,Ns1)}\period$}
\hbox{~~(the unification in the line 3 should be $\htt{Ns0}\equals_2\htt{[N2|Ns1]}$)}
\end{FRAME}

The algorithm shown in Sect.~\ref{Identification} computes three
independenet minimal inconsistent subsets; two on modes and one on
types.  Here, we do not consider Detection Rule 2 (though it can detect
the variable $\htt{Ns0}$ in the clause head of $R_2$ as an error).\\

\noindent
{\bf Minimal inconsistent subset 1 (on modes):}
\begin{center}
\begin{tabular}{r@{ }l@{ }l@{\quad}c@{\quad}c}
\hline
   \multicolumn{3}{c}{Mode constraint}& Rule& Source symbol\\
\hline
$m(\pos{\unify_1}{2})$&$=$&$\inmode$&
         \BF &``$\htt{[]}$'' in $R_1$\\
$m/\pos{\unify_1}{1}$&$=$&$m/\pos{\htt{fib}}{4}$&
         \BV &$\htt{Ns0}$ in $R_1$ \\
$m/\pos{\unify_1}{2}$&$=$&$\overline{m/\pos{\unify_1}{1}}$&
         \BU &$\unify_1$ in $R_1$\\
$m(\pos{\htt{fib}}{4})$&$=$&$\INmode$&
         \BV & $\htt{Ns0}$ in $R_2$\\
\hline
\end{tabular}
\end{center}

\noindent
{\bf Minimal inconsistent subset 2 (on modes):}
\begin{center}
\begin{tabular}{r@{ }l@{ }l@{\quad}c@{\quad}c}
\hline
   \multicolumn{3}{c}{Mode constraint}& Rule& Source symbol\\
\hline
$m(\pos{\unify_2}{2})$&$=$&$\inmode$&
         \BF &``$\htt{.}$'' in $R_2$\\
$m/\pos{\unify_2}{2}$&$=$&$\overline{m/\pos{\unify_2}{1}}$&
         \BU &$\unify_2$ in $R_2$\\
$m(\pos{\unify_2}{1})$&$=$&$\INmode$&
         \BV & $\htt{N1}$ in $R_2$\\
\hline
\end{tabular}
\end{center}

\clearpage
\noindent
{\bf Minimal inconsistent subset 3 (on types):}
\begin{center}
\begin{tabular}{r@{ }l@{ }l@{ }c@{ }c}
\hline
   \multicolumn{3}{c}{Type constraint}& Rule& Source symbol\\
\hline
$\tau/\pos{\htt{fib}}{2}$&$=$&$\tau/\pos{\htt{:=}}{2}\pos{\htt{+}}{1}$&
         (${\rm HBV}_\tau$) &$\htt{N1}$ in $R_2$\\
$\tau(\pos{\unify_2}{2})$&$=$& list type&
         (${\rm HBF}_\tau$) &``$\htt{.}$'' in $R_2$\\
$\tau/\pos{\htt{fib}}{2}$&$=$&$\tau/\pos{\unify_2}{1}$&
         (${\rm HBV}_\tau$) &$\htt{N1}$ in $R_2$\\
$\tau/\pos{\unify_2}{2}$&$=$&$\tau/\pos{\unify_2}{1}$&
         (${\rm BU}_\tau$) &$\unify_2$ in $R_2$\\
$\tau(\pos{\htt{:=}}{2}\pos{\htt{+}}{1})$&$=$& integer type&
         builtin & $\htt{:=}$ in $R_2$\\
\hline
\end{tabular}
\end{center}

These three subsets are classified into the same group because all of
the subsets indicate the clause $R_2$.
Suspected variable symbols are extracted as in the table below:

\begin{center}
\begin{tabular}{c@{~~}c@{~~}c}
\hline
Clause & Variable symbol& Subset number\\
\hline\noalign{\vskip2pt}
$R_1$ & $\htt{Ns0}$ & 1\\
$R_2$ & $\htt{Ns0}$ & 1\\
$R_2$ & $\htt{N1}$  & 2, 3\\ \hline
\end{tabular}
\end{center}

When depth-1 search is attempted, Quick-check detects that rewritings
which increase or decrease the number of occurrences of $\htt{Ns0}$ in
$R_1$ need not be considered, because such changes may dissolve the
subset 1 but neither the subset 2 nor 3.
After all, the system finds that the only possible ways to dissolve all
inconsistencies are either replacing $\htt{Ns0}$ by $\htt{N1}$ or vice
versa in $R_2$.  As the number of occurrences of $\htt{Ns0}$ and
$\htt{N1}$ is four in total, only four ways of rewriting each variable
occurrence need mode and type analyses.  Without Quick-check, a great
number of mode and type analyses would have to be done.
In this example, Kima finally finds only one alternative, which is the
program we have intended.

\section{Related Work}

Analysis of malfunctioning systems based on their intended logical
specification has been studied in the field of artificial intelligence
\cite{JAI} and known as model-based diagnosis, which has some
similarities with our work.
However, the purpose of model-based diagnosis is to analyze the
differences between intended and observed behaviors, while our system
does {\em not\/} require that the intended behavior of a program be
given as declarations.

Wand proposed an algorithm for diagnosing non-well-typed functional
programs \cite{POPL86}.  His approach was to extend the unification
algorithm for type reconstruction to record which symbol occurrence
imposed which constraint.  In contrast, our framework is built outside
any underlying framework of constraint solving.  It does not incur any
overhead for well-moded/typed programs or modify the constraint-solving
algorithm.

Tenma's system automatically corrects procedural programs under strong
typing \cite{Tenma90-e}.  When a change is made on a certain software
component, the system automatically replaces the components that do not
adapt to the change by alternative components.
%
Thus the purpose of the system is very different from Kima.  Kima works
in the situation where the locations of errors are entirely unknown, and
it works at the program symbol level rather than the software component
level.

\section{Conclusions and Future Work}
\label{future-work}

We have implemented Kima, a system which automatically corrects
near-misses in a concurrent logic program.  Kima does not have to be
given explicit specifications of program properties.

Experiments showed that, in most cases, one or a few alternatives could
be obtained from KL1 programs with a few wrong variable
occurrences. This is indebted to theoretically or statistically endorsed
heuristics as well as mode and type information.  In the set of programs
with a few mutated variable occurrences, programs that are both
well-moded/typed and with higher priority turn out to be quite rare.
Heuristic Rules and Detection Rules do not only improve detection power
and the quality of alternatives but also optimizes the search of
alternatives.

Specifications or declarations of program properties, if available, will
achieve more advanced error correction.  Our future plan is to let Kima
accept instances of a pair of input and output constraints.  Such
instances play the role of mode and type specifications also.
%

Kima is itself written in KL1 language, and is now 4,500 lines long.
The computation of minimal inconsistent subsets and the following
depth-1 search for a program of 100 lines long is completed within
several seconds.  The example shown in Sect.~\ref{exam-fibonacci} took
not more than 0.1 seconds on the KLIC system running on Sun Ultra 30
(248 MHz) + 128 MB of memory.

\newcommand{\noopsort}[1]{} \newcommand{\printfirst}[2]{#1}
  \newcommand{\singleletter}[1]{#1} \newcommand{\switchargs}[2]{#2#1}

\clearpage
\appendix

\section*{Appendix: Usage of Kima}

\subsection*{Example 1 -- A single error}

First, consider a list concatenation (append) program with one error:

\medskip
\begin{FRAME}
\hbox{~~~~~~~$\means \htt{module test}\period$}
\hbox{$R_1$ : $\htt{append([],~~~~~Y,Z~~)}\means \htt{true}\commit
 \htt{Y} \equals \htt{Z}\period$}
\hbox{$R_2$ : $\htt{append([A|Y],Y,Z0)}\means \htt{true}\commit
 \htt{Z0}\equals \htt{[A|Z]} \htt{,} \htt{append(X,Y,Z)}\period$}
\hbox{(The head of $R_2$ should have been $\htt{append([A|X],Y,Z0)}$)}
\end{FRAME}

\medskip
Suppose you want to obtain alternatives with up to priority 100 (i.e.,
very low priority), command line options should be given as:

\medskip
\begin{FRAME}
\begin{verbatim}
  % kima +p 100 append.kl1
\end{verbatim}
\end{FRAME}

\medskip
Then, Kima presents six alternatives, all up to priority 3:

\medskip
\begin{FRAME}
\small \advance\baselineskip by-3pt
\begin{verbatim}
    ================= Suspected Group 1 =================
           ------------- Priority 1 -------------
  append([A|X],Y,Z0):-true|Z0=[A|Z],append(X,Y,Z)
                          in test:append/3, clause No.2
           -----
  append([A|Y],X,Z0):-true|Z0=[A|Z],append(X,Y,Z)
                          in test:append/3, clause No.2
           -----
           ------------- Priority 2 -------------
  append([A|Y],Y,Z0):-true|Z0=[A|Z],append(Z0,Y,Z)
                          in test:append/3, clause No.2
           -----
           ------------- Priority 3 -------------
  append([A|Y],Y,Z0):-true|Z0=[A|Z],append(Y,Y,Z)
                          in test:append/3, clause No.2
           -----
  append([A|Y],Y,Z0):-true|Z0=[A|Z],append(A,Y,Z)
                          in test:append/3, clause No.2
           -----
  append([A|Y],Y,Z0):-true|Z0=[A|Z],append(Z,Y,Z)
                          in test:append/3, clause No.2
           -----
\end{verbatim}
\end{FRAME}

\medskip
The two alternatives of priority 1 have the highest priority.  
Each alternative is separeted by ``$\htt{-----}$''.
The first of the two alternatives with priority 1 is the intended one,
while the second alternative turns out to be a program that merges two
input lists by taking their elements alternately.  That is, when
$\htt{append}$ is invoked as $\htt{append([1,2,3],[4,5,6],Out)}$, the
first alternative returns $\htt{[1,2,3,4,5,6]}$ and the second returns
$\htt{[1,4,2,5,3,6]}$.

Next, let us compute minimal inconsistent subsets ({\em MIS}\/ for short)
and variable symbol occurrences infringing Detection Rules.

\medskip
\begin{FRAME}
\small \advance\baselineskip by-3pt
\begin{verbatim}
  % kima +mis append.kl1
  < Minimal Inconsistent Subsets of *Mode* constraints >
  m/<(test:append)/3,1><cons,2> = IN
          imposed by the rule HV applied to the variable Y
          in test:append/3, clause No.2
  m/<(test:append)/3,1> = OUT
          imposed by the rule BV applied to the variable X
          in test:append/3, clause No.2
  -----
  < Minimal Inconsistent Subsets of *Type* constraints >
   --Constraints are consistent, and there is no MIS--

  < Violations of syntactic rules of the detection level 2 >
  singleton(X)
          in test:append/3, clause No.2
  -----
\end{verbatim}
\end{FRAME}

\medskip 
MISs of mode constraints are obtained first; those of types second.
Multiple independent MISs can be computed at once, and each MIS is
displayed with a separator ``$\htt{-----}$''.  In this example, only one
MIS on modes is found, while type constraints are consistent.

The MIS says that variables $\htt{X}$ and
$\htt{Y}$ in the second clause of $\htt{append}$ are suspicious.  Using
this information, Kima searches alternatives either by increasing or by
decreasing the occurrences of $\htt{X}$ and/or $\htt{Y}$ in the clause.
In addition to MISs, the variable $\htt{X}$ is detected as an error
by Detection Rule 2.  Violations of Detection Rules are reported
as follows:
\begin{description}
 \item[Detection Rule 1.] \hspace{10cm}
\begin{enumerate}
 \item A variable which occurs in a clause guard must occur also in the
       head of the clause: $\htt{var\_not\_in\_the\_head({\it the variable})}$
 \item The same variable must not occur on both sides of a unification
       body goal: $\htt{not\_pass\_occur\_check({\it the variable})}$
\end{enumerate}
 \item[Detection Rule 2.]
The name of a singleton variable must begin with an underscore ``{\tt
	    \_}'': $\htt{singleton({\it the variable})}$
\end{description}

\subsection*{Example 2 -- Independent errors}

The second example is a program $\htt{comb({\it n},{\it r},Out)}$ that
generates the list of all length-$n$ 0-1 lists that contain exactly $r$
1's. (Hence the outer list contains ${}_n{\rm C}_r$ elements.)  For
example, $\htt{comb(3,2,Out)}$ returns the list\\
$\htt{[[1,1,0],[1,0,1],[0,1,1]]}$.  Below is the definition of
$\htt{comb}$ with two errors:

\medskip
\begin{FRAME}
\hbox{~~~~~~$\means \htt{module probability}\period$}
\hbox{$R_1$: $\htt{comb(N,0,C)}\means \htt{true}\commit
 \htt{init\_list(0,N,0,[],C0)} $\htt{,}$ \htt{C}\equals\htt{[C0]}\period$}
\hbox{$R_2$: $\htt{comb(N,N,C)}\means \htt{true}\commit
 \htt{init\_list(0,N,1,[],C0)} $\htt{,}$ \htt{C}\equals\htt{[C0]}\period$}
\hbox{$R_3$: $\htt{comb(N,R,C)}\means \htt{~N>R}\commit$}
\hbox{~~~~~~~~~~~$\htt{N1:=N-1}$\htt{,}$ \htt{R1:=R-1}$\htt{,}$ 
 \htt{comb(N1,R1,C0)} $\htt{,}$ \htt{cons\_list(1,C0,CC0)}$\htt{,}$$}
\hbox{~~~~~~~~~~~$\htt{comb(N1,R,C1)} $\htt{,}$ \htt{cons\_list(0,C1,CC1)}$\htt{,}$ \htt{append(CC0,CC1,CC)}\period$}
\hbox{~(The last invocation should have been $\htt{append(CC0,CC1,C)}$)}
\hbox{$R_4$: $\htt{init\_list(N,Len,\_,L0,L)}\means \htt{N=:=Len}\commit
 \htt{L0}\equals\htt{L}\period$}
\hbox{$R_5$: $\htt{init\_list(N,Len,E,L0,L)}\means \htt{N~<~Len~}\commit$}
\hbox{~~~~~~~~~~~$\htt{L1}\equals\htt{[E|L0]}$\htt{,}$ \htt{N1:=N+1}$\htt{,}$
 \htt{init\_list(N1,Len,E,L1,L)}\period$}
\hbox{$R_6$: $\htt{cons\_list(\_,[],~~~~~~~L)}\means \htt{true}\commit \htt{L}\equals\htt{[]}\period$}
\hbox{$R_7$: $\htt{cons\_list(A,[X|Xs],L)}\means \htt{true}\commit$}
\hbox{~~~~~~~~~~~$\htt{L}\equals\htt{[[A|X]|L1]}$\htt{,}$ \htt{cons\_list(A,XS,L1)}\period$}
\hbox{~(The recursive call should have been $\htt{cons\_list(A,Xs,L1)}$)}
\hbox{$R_8$: $\htt{append([],~~~~~Y,Z~~)}\means \htt{true}\commit
 \htt{Y} \equals \htt{Z}\period$}
\hbox{$R_9$: $\htt{append([A|X],Y,Z0)}\means \htt{true}\commit
 \htt{Z0}\equals \htt{[A|Z]} $\htt{,}$ \htt{append(X,Y,Z)}\period$}
\end{FRAME}

\medskip
The default action of Kima is to perform depth-1 search of alternatives
of the highest priority using modes, types, and Detection Rules.

\medskip
\begin{FRAME}
\small \advance\baselineskip by-3pt
\begin{verbatim}
  % kima comb.kl1
   ================= Suspected Group 1 =================
          ------------- Priority 1 -------------
  comb(N,R,C):-N>R|
  N1:=N-1,R1:=R-1,comb(N1,R1,C0),cons_list(1,C0,CC0),
  comb(N1,R,C1),cons_list(0,C1,CC1),append(CC0,CC1,C)
                         in probability:comb/3, clause No.3
          -----
   ================= Suspected Group 2 =================
          ------------- Priority 1 -------------
  cons_list(A,[X|Xs],L):-true|L=[[A|X]|L1],cons_list(A,Xs,L1)
                         in probability:cons_list/3, clause No.2
          -----
\end{verbatim}
\end{FRAME}

\medskip
There are two Suspected Groups.  In this example, Kima first found
multiple MISs.  By analyzing the clauses indicated by the MISs, Kima
concluded there were two independent groups.  Kima performed depth-1
search of alternatives for each group, and finally succeeded in finding
alternatives that really corrected the errors.

\subsection*{Example 3 -- Multiple errors in the same group}

Last, we consider a quicksort program with two errors in the same
clause.

\medskip
\begin{FRAME}
\hbox{~~~~~~~$\means \htt{module main}\period$}
\hbox{$R_1$ : $\htt{quicksort(Xs,Ys)}\means \htt{true}\commit
 \htt{qsort(Xs,Ys,[])}\period$}
\hbox{$R_2$ : $\htt{qsort([],~~~~~~~Ys0,Ys~~)}\means \htt{true}\commit
 \htt{Ys}\equals \htt{Ys0}\period$}
\hbox{$R_3$ : $\htt{qsort([X|Xs],Ys0,Ys3)}\means \htt{true}\commit$}
\hbox{~~~~~~~~~~~~$\htt{part(X,Xs,S,L)}\htt{,} \htt{qsort(S,Ys0,Ys1)}\htt{,}$}
\hbox{~~~~~~~~~~~~$\htt{Ys2}\equals\htt{[X|Ys1]} \htt{,} \htt{qsort(L,Ys2,Ys3)}\period$}
\hbox{~(The body unification goal should have been $\htt{Ys1}\equals\htt{[X|Ys2]}$)}
\hbox{$R_4$ : $\htt{part(\_,[],~~~~~~~S,~~L~)}\means \htt{true}\commit
 \htt{S}\equals\htt{[]} \htt{,} \htt{L}\equals\htt{[]} \period$}
\hbox{$R_5$ : $\htt{part(A,[X|Xs],S0,L~)}\means \htt{A>=X}\commit
 \htt{S0}\equals\htt{[X|S]} \htt{,} \htt{part(A,Xs,S,L)} \period$}
\hbox{$R_6$ : $\htt{part(A,[X|Xs],S,~L0)}\means \htt{A~<~X}\commit
 \htt{L0}\equals\htt{[X|L]} \htt{,} \htt{part(A,Xs,S,L)} \period$}
\end{FRAME}

\medskip
Depth-1 search is tried first, but no solution can be found.

\medskip
\begin{FRAME}
\small
\begin{verbatim}
  % kima qsort.kl1
    ================= Suspected Group 1 =================
               Sorry, no alternative is found
\end{verbatim}
\end{FRAME}

\medskip
Now depth-2 search is tried.

\medskip
\begin{FRAME}
\small
\begin{verbatim}
  % kima +d 2 qsort.kl1
    ================= Suspected Group 1 =================
           ------------- Priority 1 -------------
  qsort([X|Xs],Ys0,Ys3):-true|part(X,Xs,S,L),qsort(S,Ys0,Ys1),
  Ys1=[X|Ys2],qsort(L,Ys2,Ys3)
                          in main:qsort/3, clause No.2
           -----
\end{verbatim}
\end{FRAME}

\medskip
Only one alternative is found, and this is the intended one. In depth-2
search, depth-1 search is also executed, and all the alternatives found
by depth-1 and depth-2 searches are prioritized together. In general,
depth-$N$ search includes depth-$k$ search for all $k\leq N$.


\end{document}